\documentclass[10pt,journal,letterpaper,twoside]{IEEEtran}

\usepackage{amsmath,graphicx,cite,dsfont,amssymb}
\usepackage{amsfonts,multirow,bm,array,setspace}
\usepackage{textcomp}
\usepackage{color}
\usepackage[linesnumbered,ruled]{algorithm2e}

\newtheorem{Thm}{Theorem}

\title{Power Control for Massive MIMO Systems with Nonorthogonal Pilots}
\author{\IEEEauthorblockN{Xihan Chen, \IEEEmembership{Student Member,~IEEE}, Kaiming Shen, \IEEEmembership{Student Member,~IEEE}, Hei Victor Cheng, \IEEEmembership{Member,~IEEE},\\
An Liu, \IEEEmembership{Senior Member,~IEEE}, Wei Yu, \IEEEmembership{Fellow,~IEEE}
and Min-Jian Zhao, \IEEEmembership{Member,~IEEE}} 
\thanks{Manuscript received October 7, 2019; revised November 16, 2019. This work was supported in part by the Science and Technology Program of
Shenzhen, China, under Grant JCYJ20170818113908577, in part by the National
Natural Science Foundation of China under Project No. 61571383, and in part by Natural Sciences and Engineering Research Council (NSERC) of Canada. The work
of A. Liu was supported by the China Recruitment Program of Global Young
Experts. \emph{(Corresponding author: Kaiming Shen.)}

X. Chen, A. Liu, and M.-J. Zhao are with the College of
Information Science and Electronic Engineering, Zhejiang University,
Hangzhou 310027, China (e-mail: \{chenxihan,anliu,mjzhao\}@zju.edu.cn).

K. Shen, H. V. Cheng, and W. Yu are with
The Edward S. \mbox{Rogers} Sr. Department of Electrical and Computer
Engineering, University of Toronto, Toronto, ON M5S 3G4, Canada
(e-mail: {kshen@ece.utoronto.ca}, hei.cheng@utoronto.ca, weiyu@ece.utoronto.ca).
}}
\begin{document}


\maketitle

\begin{abstract}
This letter shows that optimizing the transmit powers along with optimally designed nonorthogonal pilots can significantly reduce pilot contamination and improve the overall throughput of the uplink multi-cell massive multiple-input multiple-output (MIMO) system as compared to the conventional schemes that use orthogonal pilots.
Given the optimized nonorthogonal pilots, power control as a function of the large-scale path-loss can be thought of as a stochastic optimization problem due to the presence of fast fading. This paper advocates a deterministic approach to solve this problem, then further proposes a stochastic optimization method that utilizes successive convex approximation as a benchmark to quantify the performance of the proposed approach. Simulation
results reveal significant advantage of using optimized nonorthogonal pilots together with power control to combat pilot contamination. 
\end{abstract}
\begin{IEEEkeywords}
Massive MIMO, nonorthogonal pilots, power control, stochastic optimization.
\end{IEEEkeywords}

\section{Introduction}
\label{sec:intro}

\IEEEPARstart{M}{assive} multiple-input multiple-output (MIMO) has attracted extensive research interests for its ability to enable interference-free transmission in an asymptotic sense, thereby improving throughput and energy efficiency significantly. However, the gain of the massive MIMO technique can be limited in practice because of the channel estimation error caused by pilot contamination. This letter aims to demonstrate the advantage of using nonorthogonal pilots together with power control for minimizing the pilot contamination effect.

Pilot design and power control are two central questions in this area (although the recent work \cite{Bjornson_TWC2018_Unlimited} shows that pilot contamination does not limit the capacity in some cases). Differing from the most existing works that use orthogonal pilots, the present work considers power control under a more general setup with nonorthogonal pilots. As shown in the recent works \cite{Eldar_arxiv18_CoordinatedPilot,Kai_arxiv19_Rx_Nonorthogonal}, using nonorthogonal pilots can already effectively improve upon the conventional orthogonal pilot scheme in terms of channel estimation. This letter further shows that the nonorthogonal pilot scheme leads to considerable throughput gain when coupled with power control.

This work formulates a multicell uplink power control  problem for the massive MIMO system in recognition of the fact that power control typically depends on the slow-varying path-loss thus takes place in a much larger time scale than the small-scale fading. Given the fixed path-loss, we aim to choose a set of transmit powers that maximize the ergodic rates taken over a large number of small-scale time intervals. This  power control problem for the uplink massive MIMO amounts to a stochastic optimization that is difficult to solve directly. Instead, this letter proposes a way of approximating the stochastic optimization problem in a deterministic form. The first part of the letter proposes a deterministic approximation to the stochastic optimization and accordingly devises an efficient iterative algorithm; the second part of the letter further gives an off-line stochastic optimization approach as benchmark to quantify the performance of the deterministic algorithm.

Power control for massive MIMO is traditionally designed for a single cell system without taking the pilot contamination effect into account \cite{Victor_2017tsp_powercontrol,Bjornson_tsp2013_ORA,Yang_tcom2017_pc}. Regarding the multi-cell scenario in the presence of pilot contamination, the earlier works \cite{Guo_ICC14,Chien_twc2018_pc,Chien_tcom2019_LSFD} consider power control by assuming that the orthogonal pilot scheme has been used for channel estimation. In contrast, the main contribution of this letter is to
explore the potential of using power control to further enhance the advantage of the nonorthogonal pilots over the traditional orthogonal pilots. As a further contribution, this paper justifies the proposed power control scheme by using a stochastic optimization based benchmark. Simulations show that the nonorthogonal pilot scheme followed by the proposed power control is more effective in mitigating pilot contamination than existing methods.

\emph{Notation:} We use $(\cdot)^*$ to denote the conjugate transpose, $\mathrm{vec}(\cdot)$ the vectorization, $\otimes$ the Kronecker product, and $\mathcal{CN}$ the complex Gaussian distribution. We use the bold letter to denote a collection of variables, e.g., $\bm p=[p_1,p_2,\ldots,p_n]$.

\setcounter{equation}{5}
\begin{figure*}
\begin{equation}\label{topE}
\gamma{}_{ik}\left(\boldsymbol{p},\boldsymbol{h}\right)=\frac{\|\hat{\bm{h}}_{i,ik}\|^{4}p_{ik}}{\sum_{(j,l)\neq(i,k)}\big|\hat{\bm{h}}_{i,ik}^{\ast}\bm{h}_{i,jl}\big|^{2}p_{jl}+\sigma^{2}\|\hat{\bm{h}}_{i,ik}\|^{2}+\big|\hat{\bm{h}}_{i,ik}^{\ast}\big(\bm{h}_{i,ik}-\hat{\bm{h}}_{i,ik}\big)\big|^{2}p_{ik}}
\end{equation}
\hrulefill
\end{figure*}
\setcounter{equation}{0}

\section{Problem Formulation}

\label{sec:setup}

Consider an uplink massive MIMO system with $I$ cells, where $M$-antenna base-station (BS) $i$ serves $K_i$ single-antenna user terminals in cell $i$. We use $i\in\left\{ 1,\dots,I\right\}$ to denote the index of each cell or the corresponding BS, and $(i,k)$ the index of the $k$th user in cell
$i$. Let $p_{ik}$ be the transmit power level of user $(i,k)$ under the power constraint $0\le p_{ik}\le P_{\max}$. Hence, for the power variable $\bm{p}=[p_{ik},\forall (i,k)]$, its feasible region is $\mathcal P=[0,P_{\max}]\times[0,P_{\max}]\times\ldots\times[0,P_{\max}]$.

Following the previous works \cite{Eldar_arxiv18_CoordinatedPilot,Kaiming_Icassp2019_multicellPilot},
we adopt the flat-fading channel model
\setcounter{equation}{0}
\begin{equation}
\label{chn}
\mathbf{H}_{ji}=\mathbf{G}_{ji}\mathbf{V}^{\frac{1}{2}}_{ji},
\end{equation}
where $\mathbf{H}_{ji}=[\bm{h}_{j,i1},\cdots,\bm{h}_{j,iK_i}]\in\mathbb{C}^{M\times K_i}$
is the channel matrix with $\bm{h}_{j,ik}\mathbb{\in C}^{M}$ denoting the channel from user $(i,k)$ to BS $j$, $\mathbf{G}_{ji}\in\mathbb{C}^{M\times K_i}$
is the small-scale fading coefficient matrix with i.i.d entries distributed
as $\mathcal{CN}(0,1)$, and $\mathbf{\mathbf{V}}_{ji}=\mathrm{diag}[v_{j,i1},\cdots,v_{j,iK_i}]\in\mathbb{C}^{K_i\times K_i}$
is the large-scale fading coefficient matrix. We assume that the set of  large-scale
fading coefficients $\{v_{j,ik}\}$ are known \emph{a priori} at each BS via the channel statistics.
%

In the pilot phase, each user $(i,k)$ transmits a pilot sequence $\bm{\phi}_{ik}\in\mathbb{C}^{L}$ of length $L$. This paper assumes that nonorthogonal pilots developed in \cite{Kai_arxiv19_Rx_Nonorthogonal} are transmitted in the pilot phase. Each BS $i$ aims to estimate its channels $\mathbf{H}_{ii}$ based on the received signal
\begin{equation}
\mathbf{Y}_{i}=\mathbf{H}_{ii}\mathbf{\Phi}_{i}^{T}+\sum_{j=1,j\neq i}^{I}\mathbf{H}_{ij}\mathbf{\Phi}_{j}^{T}+\mathbf{Z}_{i},\label{eq:received-signal}
\end{equation}
where $\mathbf{\Phi}_{i}=[\bm{\phi}_{i1},\cdots,\bm{\phi}_{iK_i}]\in\mathbb{C}^{L\times K_i}$
is the composite pilot matrix in cell $i$, and $\mathbf{Z}_{i}$
is the background noise matrix with each i.i.d. entry distributed as $\mathcal{CN}(0,\sigma^{2})$. To make the problem tractable, we further assume that the minimum mean-square error (MMSE) estimator is used for channel estimation at each BS. The resulting channel estimation is
\begin{equation}
\mathrm{vec}(\hat{\mathbf{H}}_{ii})=\left(\mathbf{V}_{ii}\mathbf{\Phi}_{i}^{\ast}\otimes\mathbf{I}_{M}\right)\left(\mathbf{U}_{i}\otimes\mathbf{I}_{M}\right)^{-1}\mathrm{vec}\left(\mathrm{\mathbf{Y}}_{i}\right),
\end{equation}
where $\mathbf{\mathbf{U}}_{i}=\sigma^{2}\mathbf{I}_{L}+\sum_{j=1}^{I}\mathbf{\Phi}_{j}\mathbf{V}_{ij}\mathbf{\Phi}_{j}^{\ast}.$

Subsequently, in the data transmission phase, each BS $i$ receives a superposition of data signals as
\begin{align}
\bm{y}_{i} & =\sum_{(j,l)}\sqrt{p_{jl}}\bm{h}_{i,jl}s_{jl}+\tilde{\bm z}_{i}, \end{align}
where $s_{ik}\sim\mathcal{CN}(0,1)$ is the data symbol of user $(i,k)$ and $\tilde{\bm{z}}_{ik}\sim\mathcal{CN}(\bm{0},\sigma^{2}\mathbf{I}_{M})$
is the background noise. Given the current channel realization $\bm h$, BS $i$ can use maximum ratio combining (MRC) to
obtain the following instantaneous data rate for its user $(i,k)$:
\begin{equation}
\label{rate}
R_{ik}\left(\boldsymbol{p},\boldsymbol{h}\right) = \log\left(1+\gamma{}_{ik}\left(\boldsymbol{p},\boldsymbol{h}\right)\right),
\end{equation}
where the signal-to-interference-plus-noise ratio (SINR) $\gamma{}_{ik}\left(\boldsymbol{p},\boldsymbol{h}\right)$ is defined in \eqref{topE} as displayed at the top of this page; more details can be found in \cite{Kai_arxiv19_Rx_Nonorthogonal}. To account for the small-scale fading, we take the expectation of $R_{ik}$ over $\bm{h}$. {Strictly speaking, $R_{ik}$ may not be achievable in practice because it requires the receiver to acquire the value of $\sum_{(j,l)\ne(i,k)}|\hat{\bm{h}}_{i,ik}^{\ast}\bm{h}_{i,jl}|^{2}p_{jl}$, rather it provides an upper bound to the achievable rate.}



More formally, the optimization problem over the power variable $\bm{p}$ (as function of the large-scale fading) is that of maximizing a network utility function of the \emph{ergodic rate} (i.e. expected rate) across all the users. Assuming a weight sum rate maximization formulation, the optimization problem becomes:
\setcounter{equation}{6}
\begin{equation}
\max_{\bm{p}\in\mathcal{P}}\:\sum_{(i,k)}w_{ik}\mathbb E_{\bm h}\big[R_{ik}\big],\label{eq:mainP}
\end{equation}
where $w_{ik}$ is a nonnegative rate weight reflecting the priority of user $(i,k)$. Note that the expectation is over the random and time-varying $\bm h$ over the small time scale, so the overall problem is a stochastic optimization problem. Differing from the case with orthogonal pilots, the ergodic rate with nonorthogonal pilots does not have a closed-form expression and is more difficult to optimize due to the last term in the denominator of \eqref{topE}.
This paper proposes to approximate problem \eqref{eq:mainP} into a deterministic form then to solve resulting problem using an efficient iterative algorithm. Furthermore, we propose a stochastic optimization based benchmark to quantify the performance of the proposed algorithm.
\section{Deterministic Optimization Algorithm}
In this section, we propose to solve (\ref{eq:mainP}) by approximating it in a deterministic form. The key enabler of this new method is a recent result from \cite{Kai_arxiv19_Rx_Nonorthogonal} that approximates the ergodic rate $\mathbb E_{\bm h}[R_{ik}(\bm p,\bm h)]$ in a deterministic form
\begin{equation}
\hat{R}_{ik}(\bm{p})=\log_{2}\bigg(1+\frac{a_{ik}p_{ik}}{\sum_{(j,l)}b_{ik,jl}p_{jl}+M\rho_{ik}\sigma^{2}-a_{ik}p_{ik}}\bigg),\label{eq:deterministic}
\end{equation}
where $\rho_{ik}=v_{i,ik}^{2}\ensuremath{\bm{\phi}_{ik}^{\ast}}\mathbf{\mathbf{U}}_{i}^{-1}\text{\ensuremath{\bm{\phi}_{ik}}}$,
$a_{ik}=M^{2}\rho_{ik}^{2}$,
and
$b_{ik,jl}=M\rho_{ik}v_{i,jl}+M^{2}v_{i,ik}^{2}v_{i,jl}^{2}\bm{\phi}_{ik}^{\ast}\mathbf{U}_{i}^{-1}\bm{\phi}_{jl}\bm{\phi}^{\ast}_{jl}\mathbf{U}_{i}^{-1}\bm{\phi}_{ik}$.

As shown in \cite{Kai_arxiv19_Rx_Nonorthogonal}, this approximate data rate $\hat{R}_{ik}(\bm{p})$ can be obtained from the so-called use-and-then-forget bound \cite{MassiveMIMO_book}. This approximate rate is always achievable but can be strictly lower than the ergodic rate $\mathbb E_{\bm h}[R_{ik}(\bm p,\bm h)]$.

Observe that $\hat{R}_{ik}(\bm{p})$ only depends on the large-scale fading $\{v_{j,ik}\}$, so it allows us to bypass the expectation over $\bm h$ and to devise a power control strategy based on the large-scale fading only. The resulting approximation of the weighted sum-rate maximization problem (\ref{eq:mainP}) is
\begin{equation}
\label{eq:WSRMP}
\max_{\bm{p}\in\mathcal{P}}\:\sum_{(i,k)}w_{ik}\hat{R}_{ik}(\bm{p}).
\end{equation}

Treating the fractional term $\frac{a_{ik}p_{ik}}{\sum_{(j,l)}b_{ik,jl}p_{jl}+M\rho_{ik}\sigma^{2}-a_{ik}p_{ik}}$ as a virtual SINR, we can recognize (\ref{eq:WSRMP}) as a deterministic power control problem that has been extensively studied in the existing literature. If we apply the idea of weighted minimum mean square error (WMMSE) algorithm \cite{luo_wmmse}, problem (\ref{eq:WSRMP}) can be reformulated as
\begin{equation}
\min_{\bm{p}\in\mathcal{P},\,\bm{\mu}\succeq\mathbf0}\:\sum_{(i,k)}w_{ik}\left(\mu_{ik}\eta_{ik}-\log_{2}\mu_{ik}\right),\label{eq:WMMSE}
\end{equation}
where $\mu_{ik}>0$ is an auxiliary variable introduced for each user $(i,k)$ and another new variable $\eta_{ik}$ is computed as
\begin{equation}
\label{eta}
\eta_{ik}=\sum_{(j,l)}b_{ik,jl}p_{jl}+M\rho_{ik}\sigma^{2}+1-2\sqrt{a_{ik}p_{ik}}.
\end{equation}
We propose an alternative optimization between $\bm p$ and $\bm \mu$ in (\ref{eq:WMMSE}) along with $\bm\eta$ updated by (\ref{eta}) iteratively. When $\bm p$ is fixed, $\bm\mu$ can be optimally determined by solving the first-order condition, that is
\begin{equation}
\label{mu}
\mu_{ik}^\star=\eta^{-1}_{ik}.
\end{equation}
Likewise, when $\bm \mu$ is held fixed, the optimal $\bm p$ is
\begin{equation}
\label{wmmse:p}
p^\star_{ik}=\Bigg[\Bigg({\sum_{(j,l)}w_{jl}\mu_{jl}b_{jl,ik}\Bigg)^{-2}}{w^2_{ik}\mu^2_{ik}a_{ik}}\Bigg]_{0}^{P_{\mathrm{max}}},
\end{equation}
where $[\cdot]^{P_{\max}}_0$ refers to $\max\{0,\min\{\cdot,P_{\max}\}\}$.

\begin{algorithm}[t]
\textbf{Input:} Pilots and large-scale fading\;
\Repeat{the increment on the value of the
objective function in \eqref{eq:WMMSE} is less than some threshold $\epsilon_1>0$}{
    Update $\bm\mu$ according to (\ref{mu})\;
    Update $\bm p$ according to (\ref{wmmse:p})\;
    Update $\bm \eta$ according to (\ref{eta});
}
\caption{Deterministic Power Control}
\label{alg:deterministic}
\end{algorithm}

We remark that Algorithm \ref{alg:deterministic} is different from the original WMMSE algorithm \cite{luo_wmmse} in that its ``SINR'' term has the desired power variable $p_{ik}$ appearing in both the numerator and the denominator. 
{
\begin{Thm}
\label{thm:deterministic}
Algorithm \ref{alg:deterministic} is guaranteed to converge to a stationary point of problem (\ref{eq:WSRMP}), with the weighted sum rate in (\ref{eq:WSRMP}) nondecreasing after each iteration.
\end{Thm}
}
The proof of this theorem is relegated to Appendix A. {Algorithm 1 is similar to the conventional power control for massive MIMO systems, except that nonorthogonal pilots are used here. The main contribution of this letter is in justifying this deterministic approximation using a stochastic optimization framework.}

\section{Stochastic Optimization Benchmark}

In order to investigate the performance of the deterministic approach (i.e., Algorithm 1), this section proposes a stochastic optimization formulation for solving problem (\ref{eq:mainP}) assuming that the successive realizations of $\bm{h}$ are observed over time. Although this algorithm is not applicable in practice, because $\bm{h}$ is never known exactly, it can still be used as a benchmark for justifying the deterministic approximation in Algorithm \ref{alg:deterministic}.

We optimize the power variable $\bm p$ in an iterative fashion, now as function of the instantaneous channel realization. Here, superscript $t$ is used to denote variables associated with the $t$th iteration. In each iteration, one realization of the channel $\{\bm{h}^t\}$ is observed. Then, instead of directly maximizing the average weighted sum rate objective, we construct a surrogate function of the objective based on the observed channel $\bm{h}^t$ to enable a successive convex approximation (SCA) \cite{Yang_TSP2016_SSCA} of the original nonconvex problem as
\begin{multline}
\label{eq:SurrogateFunction}
\hat{g}^{t}(\bm{p}^t) \!=\!\!\sum_{(i,k)}\!\Big( \alpha^{t}w_{ik}R_{ik}(\bm{p}^{t-1})+p^t_{ik}\xi_{ik}^{t}\,-\\
\frac{\tau_{ik}}{2}\big(p^t_{ik}-p_{ik}^{t-1}\big)^{2}\Big), \end{multline}
where $\alpha^t$ is the first trade-off sequence to be properly chosen, and $\tau_{ik}>0$ is an arbitrary positive constant. This function $\hat g^t(\bm{p}^t)$ is meant to approximate the original objective in (\ref{eq:mainP}); it contains an auxiliary variable iteratively updated as
\begin{equation}\label{recursiveXi}
\xi_{ik}^{t} =\alpha^{t}\sum_{(j,l)}w_{jl}\cdot\frac{\partial R_{jl}(\bm p^{t},\bm h^t)}{\partial p^t_{ik}}+\left(1-\alpha^{t}\right)\xi_{ik}^{t-1},
\end{equation}
with $\xi_{ik}^{0}=0$.
The key observation is that the new objective function $\hat g^t(\bm{p}^t)$ can be decoupled on the per-user basis, i.e.,
$\hat g^t(\bm{p}^t) = \sum_{(i,k)} q^{t}_{ik}(p^t_{ik}),$
where
\begin{multline}
\label{func_q}
q^t_{ik}(p^t_{ik}) = \alpha^{t}w_{ik}R_{ik}(\bm{p}^{t-1})+p^t_{ik}\xi_{ik}^{t}\,-\\
\frac{\tau_{ik}}{2}\big(p^t_{ik}-p_{ik}^{t-1}\big)^{2}.
\end{multline}
Thus, finding the optimal $\bm p^\star$ that maximizes the new objective $\hat g^t(\bm {p}^t)$ amounts to solving a set of separate subproblems:
\begin{equation}
\label{eq:powerP}
\mathrm{\underset{0\le\mathit{p^t_{ik}\le P_{\max}}}{\text{max}}\:}q_{ik}^{t}(p^t_{ik}).
\end{equation}
%
%
The objective function in \eqref{eq:powerP} can be recognized as
\begin{equation}
q^t_{ik}(p^t_{ik})=A (p^t_{ik})^2+Bp^t_{ik}+C
\end{equation}
for some constants $A<0$, $B$, and $C$, so it is a concave quadratic function of variable $p^t_{ik}$. As such, we can apply the first-order optimal condition to obtain the solution $\bm p^\star$, and update
$p^t_{ik}$ as

\begin{equation}
\bm p^{t}=\left(1-\beta^{t}\right)\bm p^{t-1}+\beta^{t}\bm p^\star,\label{eq:updateP}
\end{equation}
where $\beta^{t}$ is a second trade-off factor in addition to the previous $\alpha^t$; the choice of these two factors is specified later in Proposition \ref{prop:convergence}. The above steps are performed iteratively, as summarized in Algorithm \ref{alg:stochastic}.

\begin{algorithm}[t]
\textbf{Input:} Random channel samples $\bm h^t$, $t=1,2,\ldots$\;
\textbf{Initialization:} $\bm p^0\in\mathcal P$, $\theta^0_{ik}=0$, $t=0$\;
\Repeat{the increment on the value of the
objective function in \eqref{eq:SurrogateFunction} is less than some threshold $\epsilon_2>0$}{
    $t\leftarrow t+1$\;
    Construct the function $q^t_{ik}(p_{ik})$ according to (\ref{func_q})\;
    Obtain $\bm p^\star$ by solving the convex problem (\ref{eq:powerP})\;
    Compute $\bm p^{t}$ according to (\ref{eq:updateP});
}
\caption{Stochastic Optimization Benchmark}
\label{alg:stochastic}
\end{algorithm}

\begin{figure*}[t]
\begin{minipage}[b]{0.34\linewidth}
  \centerline{\includegraphics[width=6.4cm]{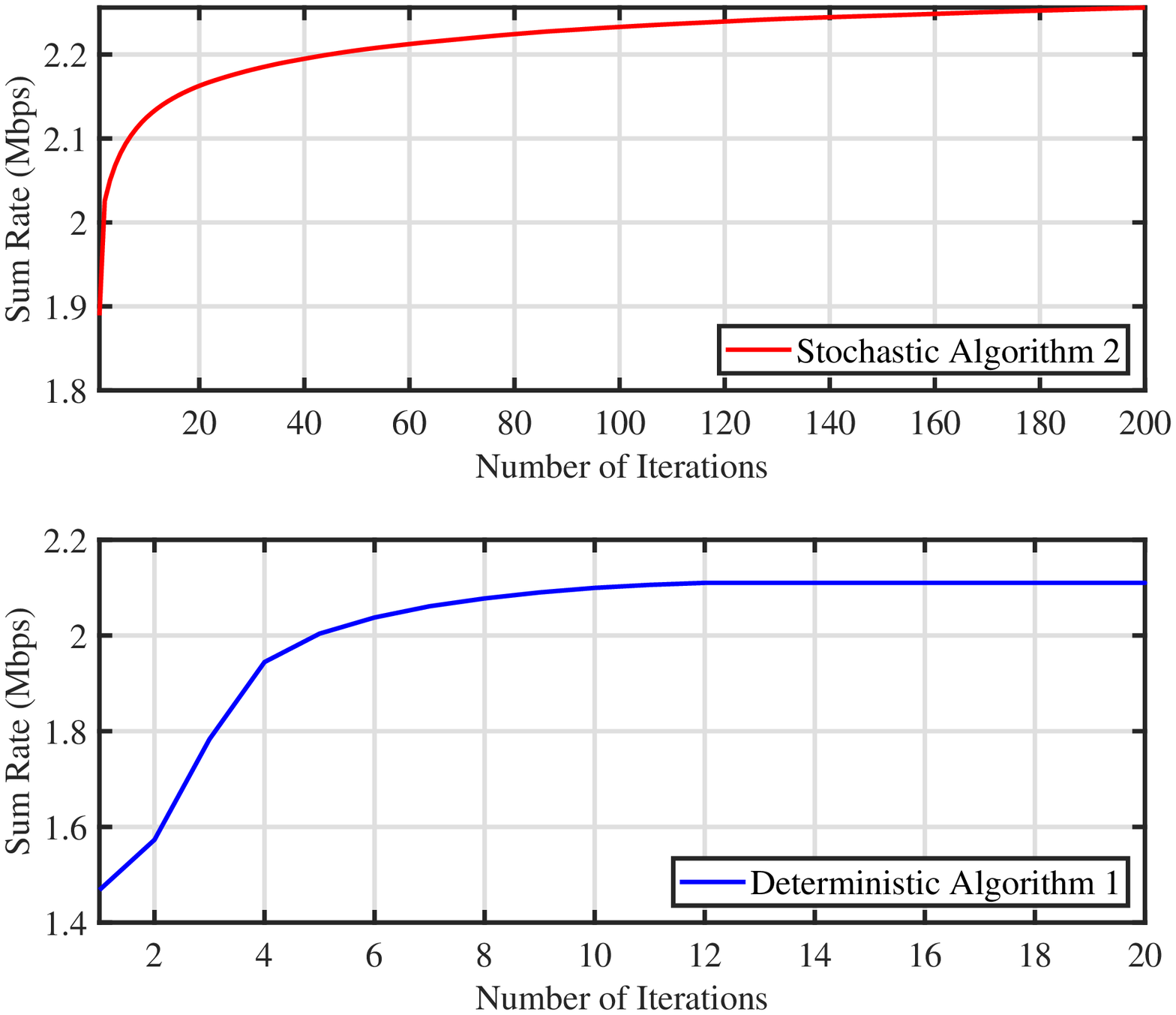}}
  \caption{Convergence of the proposed algorithms.}
  \label{fig:Convergence}
\end{minipage}
\begin{minipage}[b]{0.34\linewidth}
  \centering
  \centerline{\includegraphics[width=6.5cm]{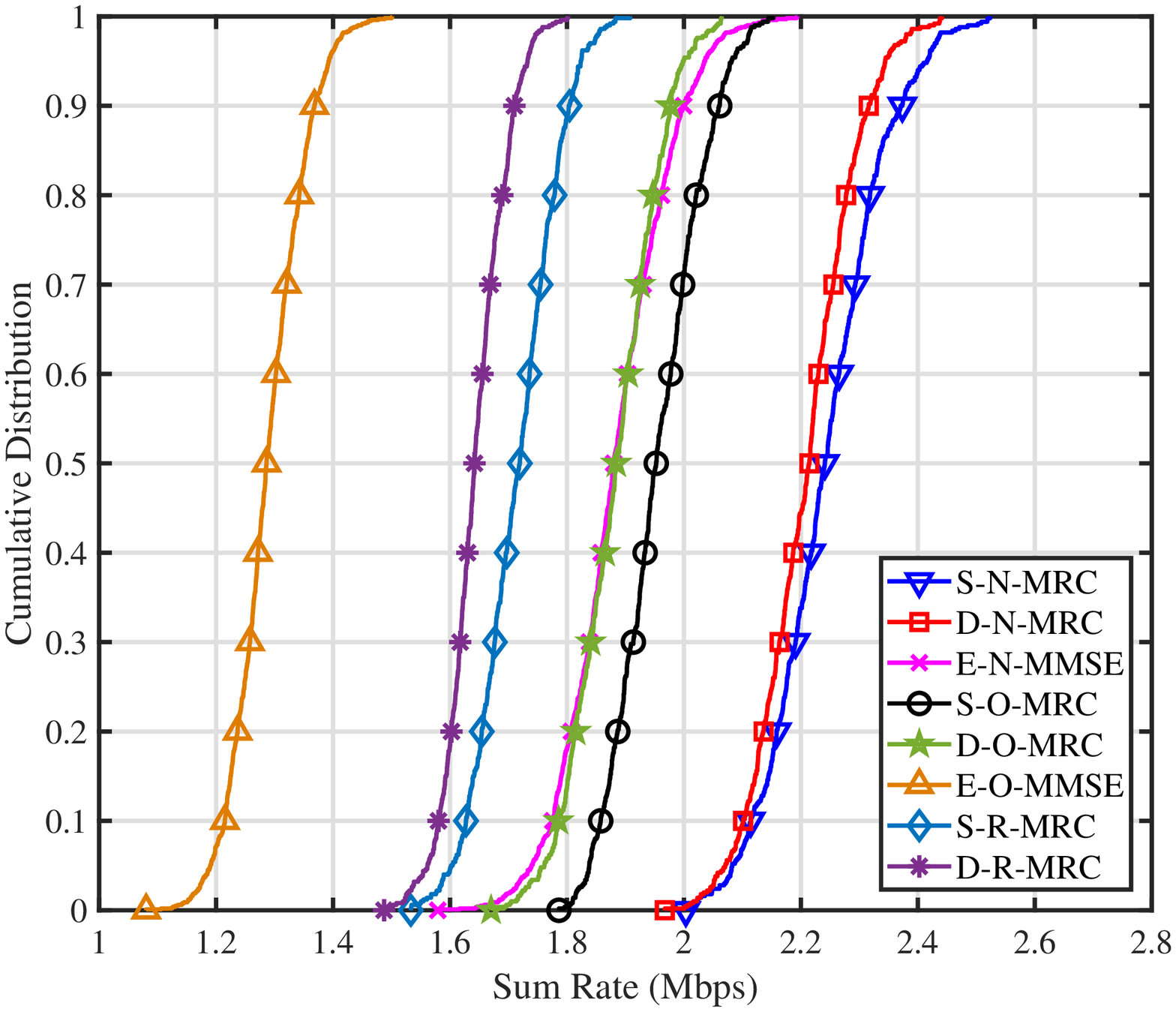}}
\caption{Cumulative distribution
function of sum rate.}
\label{fig:Cumulative-distributon-function}
\end{minipage}
\begin{minipage}[b]{0.34\linewidth}
  \centering
  \centerline{\includegraphics[width=6.5cm]{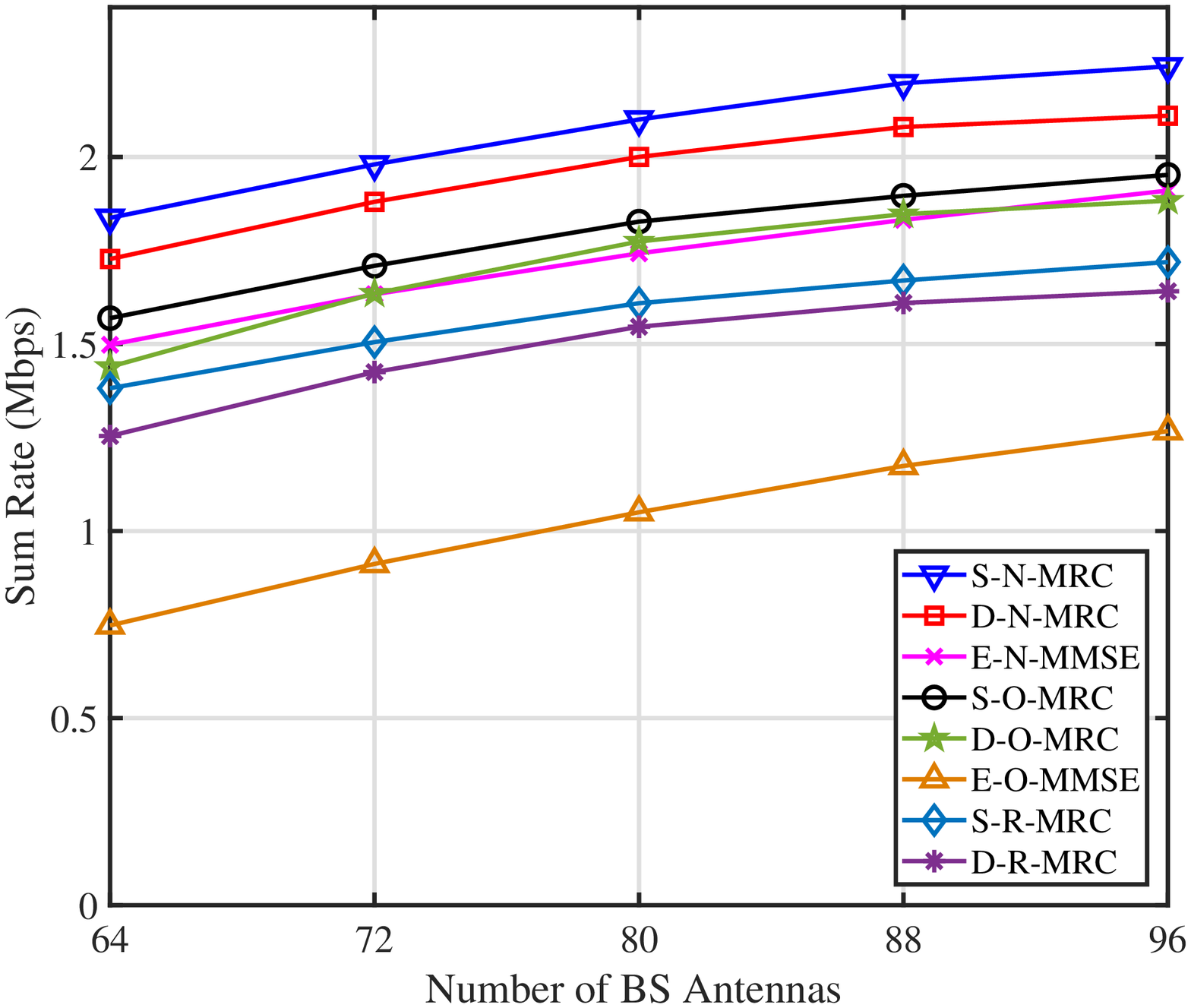}}
\caption{Sum rate vs. number of antennas at each BS.}
\label{fig:numAtn}
\end{minipage}
\end{figure*}

Algorithm \ref{alg:stochastic} can be viewed as a training process in which the power control strategy is adapted to a sequence of channel samples $\bm h^t$ generated according to (\ref{chn}). Because of this training process, Algorithm \ref{alg:stochastic} is more complex than Algorithm \ref{alg:deterministic}. However, a key advantage of Algorithm \ref{alg:stochastic} is that it guarantees convergence to a stationary point of problem (\ref{eq:mainP}) provided that the parameters $\{\alpha^t,\beta^t\}$ are chosen properly, as stated below. Thus, Algorithm \ref{alg:stochastic} can be used a benchmark to quantify the performance of Algorithm \ref{alg:deterministic}.
\begin{Thm}
\label{prop:convergence}
If the trade-off factors $\{\alpha^{t}\}$ and $\{\beta^{t}\}$ satisfy the following four conditions \cite{Yang_TSP2016_SSCA}:
\end{Thm}

\begin{enumerate}
\item $\alpha^{t}\rightarrow0$, $\frac{1}{\alpha^{t}}\leq O\left(t^{\kappa}\right)$
for some $\kappa\in\left(0,1\right)$, $\sum_{t}\left(\alpha^{t}\right)^{2}<\infty,$
\item $\beta^{t}\rightarrow0$, $\sum_{t}\beta^{t}=\infty$, $\sum_{t}\big(\beta^{t}\big)^{2}<\infty$,
\item $\lim_{t\rightarrow\infty}\beta^{t}/\alpha^{t}=0$,
\item $\underset{t\rightarrow\infty}{\lim \sup}~ \alpha^{t}\big(\sum_{(i,k)}L^t_{ik}\big)=0,$
almost surely, where $L^t_{ik}$ refers to the Lipschitz constant for the gradient of $\mathbb E_{\bm h}[R_{ik}(\bm p,\bm h)]$ with respect to $\bm p$ in the $t$-th iteration,
\end{enumerate}
then the sequence $\{\bm{p}^{t}\}$ produced by Algorithm \ref{alg:stochastic} converges to a stationary
point of problem (\ref{eq:mainP}) \emph{almost surely}. The proof of convergence is similar to that in \cite{Yang_TSP2016_SSCA}.

The motivation for some key conditions on parameters $\{\alpha^t,\beta^t\}$ is explained below. According to \eqref{eq:SurrogateFunction}, the surrogate function $\hat{g}^t(\bm{p}^{t})$ is recursively updated by averaging the instantaneous rates over a time window of size $\frac{1}{\alpha^t}$. Since $\bm{p}^t$ is changing over time $t$, the surrogate function $\hat{g}^t(\bm{p}^t)$ may not converge to  $\sum_{(i,k)}w_{ik}\mathbb E_{\bm h}\big[R_{ik}\big]$ in general. However, if $\lim_{t\rightarrow\infty}\beta^{t}/\alpha^{t}=0$, it follows from \eqref{eq:updateP} that $\bm{p}^t$ is almost unchanged within the time window $\frac{1}{\alpha^t}$
for sufficiently large $t$. In other words, $\hat{g}^t(\bm{p}^{t})$ will converge to $\sum_{(i,k)}w_{ik}\mathbb E_{\bm h}\big[R_{ik}\big]$ as $t\rightarrow\infty$, which is crucial for guaranteeing the convergence of Algorithm 2 to a stationary point of problem (\ref{eq:mainP}).

{Comparing Algorithm 2 with Algorithm 1 in term of the channel state information (CSI) cost, in Algorithm 1 each BS $j$ requires only the set of large-scale fading related to either itself or its user terminals, namely $\{\mathbf V_{ji}\;\text{and}\;\mathbf V_{ij},\forall i\}$. In contrast, Algorithm 2 depends on the small-scale fading over the time instances in addition, thus requiring the CSI $\{\mathbf H_{ji}\;\text{and}\;\mathbf H_{ij},\forall i\}$ for each BS $j$. The main point of this letter is that despite the significantly less CSI requirement, the deterministic power control in Algorithm 1 already performs close to the stochastic optimization benchmark Algorithm 2, as will be verified by simulation in Section \ref{sec:sim}.}

We further analyze the computational {complexity}. Let $I$ be the total number of BSs deployed throughout the network. Assuming that each cell has the same number of users $K$, both the deterministic algorithm and the stochastic algorithm have the computational complexity $O(K^3I^3M^2)$ per iteration. However, since the stochastic algorithm requires many more iterations to converge as compared to the deterministic,
its overall computational complexity is significantly higher, as will be verified by simulations in Section \ref{sec:sim}.

\section{Simulation Results}
\label{sec:sim}
Consider a $7$-cell wrapped-around cellular topology. The cell radius is 500 meters. A total of 9 users are uniformly distributed
in each cell. Assume that each BS has 96 antennas. The power constraint is $P_{\max}=10$
dBm. The pilot sequence has 16 symbols. The spectrum bandwidth is 1 MHz. Following the setup in {\cite{Emil_EUSP2015}}, we assume that the background noise is $-169$ dBm/Hz, and that the path-loss between user $(i,k)$ and BS $j$ is modeled as $\gamma_{j,ik}=\zeta_{j,ik}/d_{j,ik}^{3}$
, where $\zeta_{j,ik}\sim\mathcal{CN}(0,\sigma_{\zeta}^{2})$ with
the standard deviation $\sigma_{\zeta}=8$ dB is an i.i.d. log-normal
Gaussian random variable, and $d_{j,ik}$ is the distance
between user $(i,k)$ and BS $j$. The parameters $(\alpha^t,\beta^t)$ follow the diminishing stepsize rule as suggested in \cite{Yang_TSP2016_SSCA}.

We simulate the following power control methods: (i) Deterministic, namely Algorithm \ref{alg:deterministic}, denoted as ``D''; (ii) Stochastic, namely Algorithm \ref{alg:stochastic}, denoted as ``S''; (iii) Equal allocation, denoted as ``E''. We also consider three different pilot design: (i) Nonorthogonal pilots as designed in \cite{Kai_arxiv19_Rx_Nonorthogonal},
denoted as ``N''; (ii) Orthogonal, denoted as ``O''; (iii) Random (according to the Gaussian distribution), denoted as ``R''. In terms of the receiver, we consider either the MRC and the MMSE receiver. A total of eight different algorithms with different receivers, different power control methods and different pilot designs are investigated. This work advocates the deterministic power control coupled with the nonorthogonal pilots.

Fig.~\ref{fig:Convergence}(a) and (b) plot the respective objective functions of the Algorithm 1 and Algorithm 2 as functions of the iteration
number. It shows that Algorithm 1 requires only around 10 iterations to converge, while Algorithm 2 requires many more iterations. Note that Algorithm 2 is used as a benchmark for Algorithm 1. 






Fig.~\ref{fig:Cumulative-distributon-function} compares the cumulative distribution of user rates using the optimized power obtained from the different algorithms. As shown in the figure, the proposed deterministic algorithm achieves almost the same performance as the stochastic optimization benchmark with only a slight rate loss in the high rate regime.
Combined with Theorem \ref{prop:convergence}, this implies that the proposed deterministic algorithm achieves close to a stationary point of the ergodic rate maximization problem (\ref{eq:mainP}). The two algorithms with nonorthogonal pilots are superior to all the methods with orthogonal pilots. For instance, they improve upon the Stochastic-Orthogonal, which is the best among the orthogonal schemes, by around 16\% at the 50th percentile point. {Observe that all the methods with random pilots perform even worse than those with orthogonal pilots under power control (i.e., except E-O-MMSE).} It also shows that the performance of E-O-MMSE and E-O-MMSE are both inferior to D-N-MRC in terms of the data rates, which demonstrates the importance of power control to the multi-cell massive MIMO network.
Thus, with aid of power control, a simple MRC even outperforms much more complex MMSE receiver.


Fig.~\ref{fig:numAtn} shows the sum rate performance versus the number of antennas at each BS. It shows that the performance is monotonically increasing with the number of antennas; the growth rate tapers off as the number of antennas increases. Again, the proposed deterministic and stochastic algorithms have similar performance. Moreover, it is seen that the two algorithms outperform all the other benchmarks for all $M$.

\section{Conclusion}

This letter explores the potential of using power control to mitigate pilot
contamination for uplink massive MIMO under nonorthogonal pilots. The main
contribution of this letter is in showing that performing power control by
maximizing a deterministic approximation of the ergodic rates is already close
to a stochastic optimization benchmark in which power control can be
hypothetically performed over the instantaneous channel realizations.  The
proposed power control method provides significant throughput improvements upon
the classic massive MIMO system with orthogonal pilots due to its ability to
better mitigate pilot contamination.

\appendices
\section{Proof of Proposition \ref{thm:deterministic}}\label{appendixA}
We first show the existence of at least one limit point. The feasible set of the variables $(\bm{\mu},\bm{p},\bm{\eta})$ is convex and compact. It can be shown that problem (10) is upper bounded over the feasible set. Thus, the sequence of iterates produced by Algorithm 1 is compact and bounded as well. Since any compact and bounded sequence must have
at least one limit point, the existence of a limit point of Algorithm 1 is guaranteed. We now prove the equivalence between (9) and (10). We let $f_o(\bm{p})=\sum_{(i,k)}w_{ik}\hat{R}_{ik}(\bm{p})$, and let $f_r(\bm{p},\bm{\eta},\bm{\mu})= -\sum_{(i,k)} w_{ik}(\mu_{ik}\eta_{ik}-\log_2\mu_{ik})$. Moreover, we use superscript $i$ to index the iteration in Algorithm 1. Thus, $\bm{\eta}^i$ is updated by (11) with $(\bm{p}^i,\bm{\mu}^i)$, and $\bm{\mu}^i$ is updated by (12) with $(\bm{p}^i,\bm{\eta}^i)$. It turns out that
\begin{align}
f_o(\bm{p}^{i+1}) &\overset{(a)}{=} f_r(\bm{p}^{i+1},\bm{\eta}^{i+1},\bm{\mu}^{i+1})\overset{(b)}{\geq} f_r(\bm{p}^{i},\bm{\eta}^{i+1},\bm{\mu}^{i+1})
\nonumber\\
&\overset{(c)}{\geq} f_r(\bm{p}^{i},\bm{\eta}^{i},\bm{\mu}^{i+1})\overset{(d)}{\geq} f_r(\bm{p}^{i},\bm{\eta}^{i},\bm{\mu}^{i})\overset{(e)}{=} f_o(\bm{p}^{i}),\nonumber
\end{align}
where (a) and (e) both follow by the equivalence between (9) and (10); (b) follows since the update of $\bm{p}$ in (13) maximizes $f_r$ when the other variables are fixed; (c) follows since the update of $\bm{\eta}$ in (11) maximizes $f_r$ when the other variables are fixed; (d) follows since the update of $\bm{\mu}$ in (12) maximizes $f_r$ when the other variables are fixed.
Hence, the deterministic power control problem (10) is nondecreasing after each iteration. Furthermore, the convergence to a stationary point can be established by using a block coordinate descent (BCD) argument as in \cite{luo_wmmse}. The proof is thus completed.
\bibliographystyle{IEEEtran}
\bibliography{liu_all}

\end{document}